\documentclass{article}
\usepackage{etex}
\usepackage{amsmath}
\usepackage{tikz}
\usepackage{tkz-graph}
\usepackage{bussproofs}
\usepackage{graphicx}
\usepackage{color}

\usetikzlibrary{arrows,automata}
\usetikzlibrary{arrows,shapes,positioning}
\usetikzlibrary{tikzmark,decorations.pathreplacing,calc}

\newtheorem{partial solution}[theorem]{Partial Solution}

\input{tcilatex}

\begin{document}

\begin{center}
{\Large A note on Je\v{r}\'{a}bek's paper }
\end{center}

{\Large `A simplified lower bound for implicational logic\medskip '}

\begin{center}
L. Gordeev, E. H. Haeusler

\textit{Universit\"{a}t T\"{u}bingen, PUC Rio de Janeiro}

l\texttt{ew.gordeew@uni-tuebingen.de}\textit{,}

\texttt{hermann@inf.puc-rio.br\medskip }
\end{center}

{\small In \cite{GH1}, \cite{GH2}, \cite{GH3}) we sketched proofs of the
equality NP = coNP = PSPACE. These results have been obtained by proof
theoretic tree-to-dag compressing techniques adapted to Prawitz's \cite
{Prawitz} Natural Deduction (ND) for implicational minimal logic with
references to Hudelmaier's cutfree sequent calculus \cite{Hud}. In this note
we comment on Je\v{r}\'{a}bek's approach \cite{Jer} that claimed to refute
our results by providing exponential lower bounds on the implicational
minimal logic (cf. \cite{Jer}: \S 5 Conclusion). This claim is wrong and
misleading, which we briefly demonstrate below (see also \cite{GH4}) . }

{\small Je\v{r}\'{a}bek's proof of the alleged exponential lower bounds in
question deals with Frege systems (a sort of sequent calculus) for minimal
logic, whereas our universal polynomial upper bounds refer to dag-like
minimal purely implicational ND enriched by multipremise repetitions (in
short: multipremise ND). Both proof systems are sound and complete, though
not equivalent modulo provability. For some minimal tautologies are provable
in our dag-like multipremise ND by appropriate ``small'' derivations that
cannot be identified as correct proofs in any of dag-like versions of Frege
systems from \cite{Jer}. Hence exponential lower bounds claimed in \cite{Jer}
are well compatible with polynomial upper bounds in our dag-like
multipremise ND. In short, there are polynomial-time dag-like proofs in our
ND formalism that have no analogy in \cite{Jer}. The rejection also applies
to dag-like ND proofs with regard to Je\v{r}\'{a}bek's ``naive'' notion of
provability determined by the requirement \emph{``all dag-like deductive
paths are closed''} (cf. Basic example, below).}

{\small Keeping this in mind let us ask the crucial question }

{\small {$\left( Q\right) $: Is PSPACE contained in NP or not? } }

{\small To prove our affirmative answer we define a proof system $T$ in
minimal logic (cf. regular extension of $\emph{NM}_{\rightarrow }^{+}$,
below) whose provability assertion ``derivation $\partial $ proves
formula $\varphi $'' has polynomial certificate (i.e. is verifiable by a TM
in time polynomial in the weight of $\partial $) and such that $T$ proves
every minimal tautology $\varphi $ by a derivation $\partial $ of the weight
polynomial in the size of $\varphi $. Our $T$ is a suitable dag-like
multipremise modification of Prawitz's ND for purely implicational minimal
logic that proves the required assertions. This $T$\ differs from
conventional dag-like ND presentations of minimal logic. The main difference
lies in its modified notion of dag-like provability that is determined
by the requirement  \emph{``all \emph{regular} deductive paths are closed''} 
for an appropriate condition of the regularity nvolved. Note that dag-like
structure of $T$ enables us to compress ``huge'' tree-like derivations
into the desired ``small'' dag-like ones. }

{\small Also note that it is unclear how to prove $\left( Q\right) $'s
alleged negative assumption should the affirmation fail. For to conclude that a
given $L\subseteq \left\{ 0,1\right\} ^{\ast }$ is not in NP it would not
suffice to prove superpolynomial lower bounds on proofs in a chosen proof
system because no explicit description of all polynomially equivalent proof
systems involved is currently known. For instance, Je\v{r}\'{a}bek's attempt to 
address Frege systems failed, as demonstrated below by our Basic example 
showing that our dag-like multipremise ND has no analogy in \cite{Jer}. 
Summing up, exponential lower bounds claimed in \cite{Jer} are irrelevant 
to our results, while Frege systems and equivalent ND from \cite{Jer} are 
useless for $\left( Q\right) $. }

{\small \textbf{Basic example} \newline
Our basic ND calculus for minimal logic, \emph{NM}$_{\rightarrow }$,
includes two inferences $\left( \rightarrow I\right) $ and $\left(
\rightarrow E\right) $ ($\rightarrow $-\emph{introduction} and $\rightarrow $%
-\emph{elimination }\footnote{{\small {\footnotesize also known as \emph{%
modus ponens}.}}}): 
\begin{equation*}
\fbox{$\left( \rightarrow I\right) :\dfrac{\beta }{\alpha \rightarrow \beta }
$}\ \ and\ \ \fbox{$\ \left( \rightarrow E\right) \,:\dfrac{\alpha \quad
\quad \alpha \rightarrow \beta }{\beta }$}.
\end{equation*}
Its multipremise version, \emph{NM}$_{\rightarrow }^{+}$, also includes 
\emph{repetition rules} $\left( R\right) _{n}$, $n\geq 1$: 
\begin{equation*}
\fbox{$\ \left( R\right) _{n}:\overset{n\ times}{\dfrac{\overbrace{\alpha
\quad \cdots \quad \alpha }}{\alpha }}\ $}
\end{equation*}
where $\alpha $, $\beta $, $\gamma $, etc. are arbitrary purely
implicational formulas of minimal logic. A given deductive path $\Theta \!=\!%
\left[ x_{0},\cdots ,x_{h}=r\right] $ connecting leaf $x_{0}$ labeled $%
\alpha $ with the root $r$ labeled $\rho $ is called \emph{closed}
(otherwise \emph{open}) if there exists $i<h$ such that $x_{i}$ is the
conclusion of $\left( \rightarrow I\right) $ labeled $\alpha \rightarrow
\beta $ (for some $\beta $); these \emph{discharged} assumptions are
decorated as $\left[ \alpha \right] $. }

{\small %\begin{example}
Consider a tree-like deduction $\partial \in \emph{NM}_{\rightarrow }$ of $%
\tau \rightarrow \sigma $:$\smallskip $ $\smallskip $ $\smallskip $ }

{\small \hspace{-20pt} 
\begin{tikzpicture}
  \draw (0.6,2.5) node {
    \AxiomC{$[\alpha]^{1}$}
    \AxiomC{$[\alpha \rightarrow\beta]^{2}$}
    \BinaryInfC{$\beta$}
    \UnaryInfC{$\gamma \rightarrow \beta$}    
    \UnaryInfC{$ (\alpha\rightarrow \beta) \rightarrow(\gamma \rightarrow \beta)^{2}$}
    \UnaryInfC{$\alpha\rightarrow ((\alpha\rightarrow \beta) \rightarrow(\gamma \rightarrow \beta))^{1}$}
    \UnaryInfC{$\alpha\rightarrow ((\alpha\rightarrow \beta) \rightarrow(\gamma \rightarrow \beta))$}
    \AxiomC{$[\delta]^{3}$}
    \AxiomC{$[\delta\rightarrow\beta]^{4}$}
    \BinaryInfC{$\beta$}
    \UnaryInfC{$\gamma \rightarrow \beta$}  
    \UnaryInfC{$(\delta\rightarrow \beta) \rightarrow(\gamma \rightarrow \beta)^{4}$}
    \UnaryInfC{$\delta \rightarrow ((\delta\rightarrow \beta) \rightarrow(\gamma \rightarrow \beta))^{3}$}
    \AxiomC{$[\tau]^{5}$}
    \UnaryInfC{$\left.\overset{} \tau \right.$}
    \UnaryInfC{$\left.\overset{}\tau\right.$}
    \UnaryInfC{$\left. \overset{}\tau \right.$}
    \UnaryInfC{$\left.\overset{} \tau \right.$}
   \BinaryInfC{$(\alpha\rightarrow ((\alpha \rightarrow \beta) \rightarrow (\gamma\rightarrow\beta))) \rightarrow \sigma$}
    \BinaryInfC{$\sigma$}
    \UnaryInfC{$\tau \rightarrow \sigma^{5}$}
   \UnaryInfC{$$}
    \DisplayProof
  } ;
\end{tikzpicture}
\newline
where \newline
$\tau :=(\delta \rightarrow((\delta \rightarrow\beta) \rightarrow (\gamma
\rightarrow \beta))) \rightarrow ((\alpha \rightarrow ((\alpha \rightarrow
\beta) \rightarrow (\gamma \rightarrow\beta))) \rightarrow \sigma)$. \newline
Obviously every deductive path in $\partial $ is closed, and hence $\tau
\rightarrow \sigma $ is provable in \emph{NM}$_{\rightarrow }$. }

{\small Define a compressed dag-like deduction $\partial ^{\prime }\in \emph{%
NM}_{\rightarrow }^{+}$ by merging two occurrences $\gamma \rightarrow \beta 
$ using the repetition $\left( R\right) _{2}$ with conclusion $\beta $ :%
\newline
}

{\small \hspace{-20pt} 
\begin{tikzpicture}
  \draw (0,2.5) node {
    \AxiomC{$[\alpha]^{1}$}
    \AxiomC{$[\alpha \rightarrow\beta]^{2}$}
   % \LeftLabel{$\left\{ \textcolor{red}{1}\right\}$}
   % \RightLabel{$\left\{ \textcolor{red}{2}\right\}$}
    \BinaryInfC{$\beta $}
    \AxiomC{$[\delta]^{3}$}
    \AxiomC{$[\delta\rightarrow\beta]^{4}$}
   %\LeftLabel{$\left\{ \textcolor{red}{3}\right\}$}
   %\RightLabel{$\left\{ \textcolor{red}{4}\right\}$}
    \BinaryInfC{$\beta$} 
   %\LeftLabel{$\left\{ \textcolor{red}{1},\textcolor{red}{2}\right\}$}
   %\RightLabel{$\left\{\textcolor{red}{3}, \textcolor{red}{4}\right\}$}
    \BinaryInfC{$\beta$}
   %\LeftLabel{$\left\{\textcolor{red}{1}, \textcolor{red}{2}, \textcolor{red}{3}, \textcolor{red}{4}\right\}$}
    \UnaryInfC{$\gamma \rightarrow \beta$} 
    %\LeftLabel{$\left\{ \textcolor{red}{1},\textcolor{red}{2}, \textcolor{red}{3}, \textcolor{red}{4}\right\}$}
    \UnaryInfC{$$} 
       \DisplayProof
  } ;
  \draw [->] (0,1.5) -- (2.5,1) ;
  \draw [->] (0,1.5) -- (-2.5,1) ;
  \draw (1,0) node {
      \AxiomC{}
   % \LeftLabel{$\left\{  \textcolor{red}{1},\textcolor{red}{2}, 3, 4\right\}$\!}\!
    \UnaryInfC{\!$ (\alpha\!\rightarrow \!\beta) \!\rightarrow\!(\gamma \!\rightarrow \!\beta)^{2}$}
   %\LeftLabel{$\left\{ \textcolor{red}{1}, \! 2, \! 3, \! 4\right\}$\!}\!
    \UnaryInfC{\!$\alpha\!\rightarrow \!((\alpha\!\rightarrow \!\beta)\! \rightarrow\!(\gamma\! \rightarrow \!\beta))^{1}$}
   %\LeftLabel{$\left\{ 1, \! 2, \! 3, \! 4\right\}$\!}\!
    \UnaryInfC{$\alpha\!\rightarrow \!((\alpha\rightarrow\! \beta)\! \rightarrow\!(\gamma\! \rightarrow\! \beta))$}
    \AxiomC{}
   %\RightLabel{$\left\{1, \! 2, \! \textcolor{red}{3}, \! \textcolor{red}{4}\right\}$}    
    \UnaryInfC{\!\!$(\delta\!\rightarrow\! \beta) \!\rightarrow\!(\gamma \!\rightarrow \!\beta)^{4}$}
   %\LeftLabel{$\!\!\left\{ 1,  \!2,  \!\textcolor{red}{3}, \!4\right\}$\!}\! 
   \UnaryInfC{$\!\!\delta\! \rightarrow\! ((\delta\!\rightarrow\! \beta)\! \rightarrow\!(\gamma \!\rightarrow \!\beta))^{3}$}
    \AxiomC{$\QDATOP{\left[ \tau \right]^{5} }{ \,\downarrow }$}
   %\RightLabel{$\left\{\textcolor{red}{5}\right\}$}
    \UnaryInfC{$\left. \overset{}\tau \right.$}
   %\RightLabel{$\left\{ \textcolor{red}{5}\right\}$}
    \UnaryInfC{$\left. \overset{%
}\tau \right.$}
   %\LeftLabel{$\left\{ 1, \! 2, \! 3, \! 4\right\}$\!}\!
   %\RightLabel{$\left\{ \textcolor{red}{5}\right\}$}
    \BinaryInfC{$(\alpha\!\rightarrow\! ((\alpha\! \rightarrow\! \beta)\! \rightarrow\! (\gamma\!\rightarrow\!\beta))) \!\rightarrow \!\sigma$}
   %\LeftLabel{$\left\{ 1, \! 2, \! 3, \! 4\right\}$\!}\!
   %\RightLabel{$\left\{1,  \!2,  \!3,  \!4,  \!\textcolor{red}{5}\right\}$}
    \BinaryInfC{$\sigma$}
   %\LeftLabel{$\left\{1,  \!2,  \!3,  \!4,  \!\textcolor{red}{5}\right\}$} 
    \UnaryInfC{$\tau \!\rightarrow \!\sigma^{5}$} 
   %\LeftLabel{$\left\{1,  \!2,  \!3,  \!4,  \!5\right\}$} 
   \UnaryInfC{$$}
   
 \DisplayProof
  } ;
\end{tikzpicture}
\newline
Note that $\partial ^{\prime } $ has open deductive paths like those
connecting assumptions $\alpha $ (or $\alpha \rightarrow \beta $) with $\tau
\rightarrow \sigma $ via $(\delta \rightarrow \beta )\rightarrow (\gamma
\rightarrow \beta )$ and/or $\delta $ (or $\delta \rightarrow \beta $) via $%
(\alpha \rightarrow \beta )\rightarrow (\gamma \rightarrow \beta )$.
However, the closure of other (called \emph{regular}) deductive paths in $%
\partial ^{\prime }$ already confirms the provability of $\tau \rightarrow
\sigma $. This is readily seen by tree-like unfolding of $\partial ^{\prime
} $ back to $\partial $ that eliminates all open dag-like deductive paths in
question. Now the regular deductive paths are naturally determined by the
repetition rule $\left( R\right) _{2}$ involved. (In the general case of
arbitrary dag-like proofs we extend \emph{NM}$_{\rightarrow } ^{+}$ by
adding appropriate conditions of regularity that enable to formalize this
notion in polynomial way). }

{\small Note that Frege systems used in \cite{Jer} are claimed equivalent to
corresponding (both tree- and dag-like verions of) ND under the provability 
\emph{``all deductive paths are closed''} that as we just demonstrated is
not a necessary condition for dag-like provability, whereas
a weaker sufficient condition \emph{``all \emph{regular} deductive paths are
closed''} enables to prove our claim NP = coNP = PSPACE. }

\end{document}